\documentclass[prd,showpacs,showkeys,nofootinbib,twocolumn]{revtex4-1}

\usepackage{graphicx}
\usepackage{amsmath,amsfonts,amssymb}

\begin{document}

\def\ut{{\underset {\widetilde{\ \ }}u}}
\def\at{{\underset {\widetilde{\ \ }}a}}
\def\ap{\check{a}}

\title{Deviation equation in Riemann-Cartan spacetime}

\author{Dirk Puetzfeld}
\email{dirk.puetzfeld@zarm.uni-bremen.de}
\homepage{http://puetzfeld.org}
\affiliation{University of Bremen, Center of Applied Space Technology and Microgravity (ZARM), 28359 Bremen, Germany} 

\author{Yuri N. Obukhov}
\email{obukhov@ibrae.ac.ru}
\affiliation{Theoretical Physics Laboratory, Nuclear Safety Institute, 
Russian Academy of Sciences, B.Tulskaya 52, 115191 Moscow, Russia} 

\date{ \today}

\begin{abstract}
We derive a generalized deviation equation in Riemann-Cartan spacetime. The equation describes the dynamics of the connecting vector which links events on two general adjacent world lines. Our result is valid for any theory in a Riemann-Cartan background, in particular, it is applicable to a large class of gravitational theories which go beyond the general relativistic framework. 
\end{abstract}


\maketitle


\section{Introduction}\label{introduction_sec}

Within the theory of General Relativity, the relative motion of test bodies is described by means of the geodesic deviation (Jacobi) equation \cite{LeviCivita:1926,Synge:1926,Synge:1927,Pirani:1956}. This equation only holds under certain assumptions and can be used only for the description of structureless neutral test bodies. 

In a previous work \cite{Puetzfeld:Obukhov:2016:1}, we have worked out generalized versions of the deviation equation, see also \cite{Plebanski:1965,Hodgkinson:1972,Bazanski:1974,Bazanski:1976,Bazanski:1977:1,Bazanski:1977:2,Novello:etal:1977,Aleksandrov:Piragas:1978,Mannoff:1979,Schattner:Truemper:1981,Swaminarayan:etal:1983,Schutz:1985,Ciufolini:1986,Bazanski:etal:1987:1,Bazanski:etal:1987:2,Bazanski:1988,Vanzo:1992,Kerner:etal:2001,Mannoff:2001,Holten:2002,Chicone:Mashhoon:2002,Perlick:2008,Vines:2014} for alternative derivations and generalizations. Our findings in \cite{Puetzfeld:Obukhov:2016:1} extended the range of applicability of the deviation equation to general world lines. However, the results were limited to theories in a Riemannian background. While such theories are justified in many physical situations, several modern gravitational theories \cite{Hehl:1976,Blagojevic:Hehl:2013,Ponomarev:Bravinsky:Obukhov:2017} reach significantly beyond the Riemannian geometrical framework. In particular, it is already well known \cite{Hehl:Obukhov:Puetzfeld:2013,Puetzfeld:Obukhov:2014:2,Obukhov:Puetzfeld:2015:1} that in the description of test bodies with intrinsic degrees of freedom -- like spin -- there is a natural coupling to the post-Riemannian features of spacetime. Therefore, in view of possible tests of gravitational theories by means of structured test bodies, a further extension of the deviation equation to post-Riemannian geometries seems to be overdue.

In this work we derive a generalized deviation equation in a Riemann-Cartan background, allowing for spacetimes endowed with torsion. This equation describes the dynamics of the connecting vector which links events on two general (adjacent) world lines. Our results are valid for any theory in a Riemann-Cartan background; in particular, they apply to Einstein-Cartan theory \cite{Trautman:2006} as well as to Poincar\'e gauge theory \cite{Obukhov:2006}. 

The structure of the paper is as follows: In section \ref{sec_world_dev}, we briefly introduce the concepts needed in the derivation of the exact generalized deviation equation. This is followed by section \ref{rc_world_function_sec}, in which we focus on the properties of a world function based on autoparallels in a Riemann-Cartan background.  These results are then applied in section \ref{sec_rc_dev} to arrive at an expanded approximate version of the deviation equation. In section \ref{sec_coordinate_choice} we discuss how different coordinate choices, depending on the underlying gravity theory, affect the interpretation and the operational value of the deviation equation. We conclude our paper in section \ref{sec_conclusions} with a discussion of the results obtained and with an outlook of their possible applications. Our notations and conventions are summarized in appendix \ref{sec_notation} and table \ref{tab_symbols}. Some details and intermediate results of our derivation are given in appendix \ref{sec_intermediate_details}.
 
\section{World function and deviation equation}\label{sec_world_dev}

Let us briefly recapitulate the relevant steps that lead to the generalized deviation equation as derived in \cite{Puetzfeld:Obukhov:2016:1}. We want to compare two general curves $Y(t)$ and $X(\tilde{t})$ in an arbitrary spacetime manifold. Here $t$ and $\tilde{t}$ are general parameters, i.e.\ not necessarily the proper time on the given curves. In contrast to the Riemannian case, we now connect two points $x\in X$ and $y\in Y$ on the two curves by the autoparallel joining the two points (we assume that this autoparallel is unique). An autoparallel is a curve along which the velocity vector is transported parallel to itself with respect to the connection on the spacetime manifold. In a Riemannian space autoparallel curves coincide with geodesic lines. 

\begin{figure}
\begin{center}
\includegraphics[width=7cm,angle=-90]{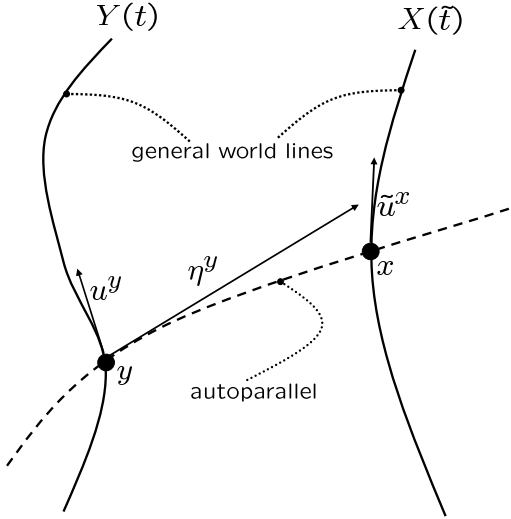}
\end{center}
\caption{\label{fig_setup} Sketch of the two arbitrarily parametrized world lines $Y(t)$ and $X(\tilde{t})$ and the (dashed) autoparallel connecting two points on these world lines. The generalized deviation vector along the reference world line $Y$ is denoted by $\eta^y$.}
\end{figure}

Along the autoparallel we have the world function $\sigma$, and conceptually the closest object to the connecting vector between the two points is the covariant derivative of the world function, denoted at the point $y$ by $\sigma^y$, cf.\ fig.\ \ref{fig_setup}. Following our conventions, the reference curve will be $Y(t)$, and we define the generalized connecting vector to be
\begin{eqnarray}
\eta^y := - \sigma^y \label{gen_dev_definition} 
\end{eqnarray}
Taking its covariant total derivative, we have
\begin{eqnarray}
\frac{D}{dt} \eta^{y_1} &=& - \sigma^{y_1}{}_{y_2} u^{y_2} - \sigma^{y_1}{}_{x_2} \tilde{u}^{x_2} \frac{d\tilde{t}}{dt}, \label{eta_1st_deriv}
\end{eqnarray}
where the velocities along the two curves $Y$ and $X$ are defined as $u^{y}:= {d Y^{y}}/{d t}$, and $\tilde{u}^{x}:={d X^{x}}/{d \tilde{t}}$. Denoting the accelerations by $a^y:={D u^y}/dt$, and $\tilde{a}^x:={D \tilde{u}^x}/d\tilde{t}$, the second derivative becomes
\begin{eqnarray}
\frac{D^2}{dt^2} \eta^{y_1} &=& - \sigma^{y_1}{}_{y_2 y_3} u^{y_2} u^{y_3} - 2 \sigma^{y_1}{}_{y_2 x_3} u^{y_2} \tilde{u}^{x_3} \frac{d\tilde{t}}{dt} \nonumber \\
&& - \sigma^{y_1}{}_{y_2} a^{y_2} - \sigma^{y_1}{}_{x_2 x_3} \tilde{u}^{x_2} \tilde{u}^{x_3} \left(\frac{d\tilde{t}}{dt} \right)^2 \nonumber \\
&& - \sigma^{y_1}{}_{x_2} \tilde{a}^{x_2} \left(\frac{d\tilde{t}}{dt} \right)^2 - \sigma^{y_1}{}_{x_2} \tilde{u}^{x_2}  \frac{d^2\tilde{t}}{dt^2}. \label{eta_2nd_deriv}
\end{eqnarray}
The generalized deviation equation is obtained from (\ref{eta_2nd_deriv}) by expressing all quantities therein along the reference word line $Y$. With the help of the inverse $\stackrel{-1}{\sigma}$ -- obtained from $\stackrel{-1}{\sigma}{}\!\!^{y_1}{}_x \sigma^{x}{}_{y_2} = \delta^{y_1}{}_{y_2}$ and $\stackrel{-1}{\sigma}{}\!\!^{x_1}{}_y \sigma^{y}{}_{x_2} = \delta^{x_1}{}_{x_2}$ -- and by defining the Jacobi propagators $H^{x_1}{}_{y_2} := - \stackrel{-1}{\sigma}{}\!\!^{x_1}{}_{y_2}$ and $K^{x_1}{}_{y_2} := - \stackrel{-1}{\sigma}{}\!\!^{x_1}{}_{y_1} \sigma^{y_1}{}_{y_2}$, the velocity along $X$ may be expressed as
\begin{eqnarray}
\tilde{u}^{x_3} &=& K^{x_3}{}_{y_2} u^{y_2} \frac{dt}{d\tilde{t}} - H^{x_3}{}_{y_1} \frac{D \sigma^{y_1}}{dt} \frac{dt}{d\tilde{t}}, \label{formal_vel_2}
\end{eqnarray} 
and inserted into (\ref{eta_2nd_deriv})
\begin{eqnarray}
\frac{D^2}{dt^2} \eta^{y_1} &=& - \sigma^{y_1}{}_{y_2 y_3} u^{y_2} u^{y_3} - \sigma^{y_1}{}_{y_2} a^{y_2} - \sigma^{y_1}{}_{x_2} \tilde{a}^{x_2} \left(\frac{d\tilde{t}}{dt} \right)^2  \nonumber \\
&&- 2 \sigma^{y_1}{}_{y_2 x_3} u^{y_2} \left( K^{x_3}{}_{y_4} u^{y_4} - H^{x_3}{}_{y_4} \frac{D \sigma^{y_4}}{dt} \right) \nonumber \\
&& - \sigma^{y_1}{}_{x_2 x_3}  \left( K^{x_2}{}_{y_4} u^{y_4} - H^{x_2}{}_{y_4} \frac{D \sigma^{y_4}}{dt} \right) \nonumber \\
&& \times \left( K^{x_3}{}_{y_5} u^{y_5} - H^{x_3}{}_{y_5} \frac{D \sigma^{y_5}}{dt} \right) \nonumber \\
&& - \sigma^{y_1}{}_{x_2} \frac{d t}{d \tilde{t}} \frac{d^2\tilde{t}}{dt^2} \left( K^{x_2}{}_{y_3} u^{y_3} - H^{x_2}{}_{y_3} \frac{D \sigma^{y_3}}{dt} \right). \label{eta_2nd_deriv_alternative_2}
\end{eqnarray}
Note that we may determine the factor $d \tilde{t} / dt $ by requiring that the velocity along the curve $X$ is normalized, i.e.\ $\tilde{u}^x \tilde{u}_x =1$, in which case we have
\begin{eqnarray}
\frac{d\tilde{t}}{dt} &=& \tilde{u}_{x_1} K^{x_1}{}_{y_2} u^{y_2} -  \tilde{u}_{x_1} H^{x_1}{}_{y_2} \frac{D \sigma^{y_2}}{dt}. \label{formal_vel_3}
\end{eqnarray} 
Equation (\ref{eta_2nd_deriv_alternative_2}) is the exact generalized deviation equation, it is completely general and can be viewed as the extension of the standard geodesic deviation (Jacobi) equation to any order. In particular, it allows for a comparison of two general, i.e.\ not necessarily geodetic or autoparallel, world lines in spacetime. 

\section{World function in Riemann-Cartan spacetime}\label{rc_world_function_sec}

In this section we work out the basic properties of a world function $\sigma$ based on autoparallels in a Riemann-Cartan spacetime, which, in contrast to a Riemannian spacetime, is endowed with an asymmetric connection $\Gamma_{ab}{}^c$. Relevant references that contain some results in a Riemann-Cartan context are \cite{Goldthorpe:1980,NiehYan:1982,Barth:1987,Yajima:1996,Manoff:2000:1,Manoff:2001:1,Iliev:2003:1,Hoogen:2017}. 

For a world function $\sigma$ based on autoparallels, we have the following basic relations in the case of spacetimes with asymmetric connections:
\begin{eqnarray}
\sigma^x \sigma_x = \sigma^y \sigma_y &=& 2 \sigma, \label{rel_1}\\
\sigma^{x_2} \sigma_{x_2}{}^{x_1} &=& \sigma^{x_1}, \label{rel_2}\\
\sigma_{x_1 x_2} - \sigma_{x_2 x_1} &=& T_{x_1 x_2}{}^{x_3} \partial_{x_3} \sigma.  \label{sigma_commutator}
\end{eqnarray}
Note, in particular, the change in (\ref{sigma_commutator}) due to the presence of the spacetime torsion $T_{x_1 x_2}{}^{x_3}$, which leads to $\sigma_{x_1 x_2} \neq \sigma_{x_2 x_1}$, in contrast to the symmetric Riemannian case, in which $\overline{\sigma}_{x_1 x_2} \stackrel{\rm s}{=} \overline{\sigma}_{x_2 x_1}$ holds.\footnote{We use ``s'' to indicate relations which only hold for symmetric connections and denote Riemannian objects by the overbar.}

In many calculations the limiting behavior of a bitensor $B_{\dots}(x,y)$ as $x$ approaches the references point $y$ is required. This so-called coincidence limit of a bitensor $B_{\dots}(x,y)$ is a tensor 
\begin{eqnarray}
\left[B_{\dots} \right] = \lim\limits_{x\rightarrow y}\,B_{\dots}(x,y),\label{coin}
\end{eqnarray}
at $y$ and will be denoted by square brackets. In particular, for a bitensor $B$ with arbitrary indices at different points (here just denoted by dots), we have the rule \cite{Synge:1960}
\begin{eqnarray}
\left[B_{\dots} \right]_{;y} = \left[B_{\dots ; y} \right] + \left[B_{\dots ; x} \right]. \label{synges_rule}
\end{eqnarray}
We collect the following useful identities for the world function $\sigma$:
\begin{eqnarray}
{}[\sigma]=[\sigma_x]=[\sigma_y]&=&0 , \label{coinc_1}\\
{}[\sigma_{x_1 x_2}]=[\sigma_{y_1 y_2}]&=&g_{y_1 y_2} , \label{coinc_2}\\
{}[\sigma_{x_1 y_2}]=[\sigma_{y_1 x_2}]&=&-g_{y_1 y_2} , \label{coinc_3}\\
{}[\sigma_{x_3 x_1 x_2}] + [\sigma_{x_2 x_1 x_3}] &=&0. \label{coinc_4}
\end{eqnarray}
Note that up to the second covariant derivative the coincidence limits of the world function match those in spacetimes with symmetric connections. However, at the next (third) order the presence of the torsion leads to
\begin{eqnarray}
{}[\sigma_{x_1 x_2 x_3}] &=& \frac{1}{2} \left(T_{y_1 y_3 y_2} + T_{y_2 y_3 y_1}+ T_{y_1 y_2 y_3} \right) = K_{y_2 y_1 y_3} ,\nonumber\\
\end{eqnarray}
where in the last line we made use of the contortion $K$, cf.\ also appendix \ref{sec_notation} for an overview of the geometrical quantities.\footnote{The contortion $K_{y_2 y_1 y_3}$ should not be confused with the Jacobi propagator $K^x{}_y$.} With the help of (\ref{synges_rule}), we obtain for the other combinations with three indices
\begin{eqnarray}
{}[\sigma_{x_1 x_2 y_3}] &=& -[\sigma_{x_1 x_2 x_3}] =  [\sigma_{y_1 x_2 y_3}] = [\sigma_{x_2 x_3 x_1}] = K_{y_2 y_3 y_1},  \nonumber\\
{}[\sigma_{x_1 y_2 y_3}] &=& -[\sigma_{x_2 x_3 x_1}] = [\sigma_{x_1 x_3 x_2}] = [\sigma_{y_1 x_2 x_3}] = K_{y_3 y_1 y_2},  \nonumber\\
{}[\sigma_{y_1 y_2 y_3}] &=& -[\sigma_{x_3 x_2 x_1}] = K_{y_2 y_1 y_3},  \nonumber\\
{}[\sigma_{x_1 y_2 x_3}] &=& -[\sigma_{x_1 x_3 x_2}] = K_{y_3 y_2 y_1},  \nonumber\\
{}[\sigma_{y_1 y_2 x_3}] &=& \phantom{-}[\sigma_{x_3 x_2 x_1}] = K_{y_2 y_3 y_1}.  \label{coinc_5}
\end{eqnarray}
The non-vanishing of these limits leads to added complexity in subsequent calculations compared to the Riemannian case. 

At the fourth order we have
\begin{eqnarray}
&& K_{y_1}{}^{y}{}_{y_2} K_{y_3 y y_4} + K_{y_1}{}^{y}{}_{y_3} K_{y_2 y y_4} + K_{y_1}{}^{y}{}_{y_4} K_{y_2 y y_4} \nonumber \\
&&+[\sigma_{x_4 x_1 x_2 x_3}] + [\sigma_{x_3 x_1 x_2 x_4}] + [\sigma_{x_2 x_1 x_3 x_4}] =0, \label{coinc_6}\end{eqnarray}
and in particular
\begin{widetext}
\begin{eqnarray}
{}[\sigma_{x_1 x_2 x_3 x_4}] &=&\frac{1}{3}  \nabla_{y_1} \left( K_{y_3 y_2 y_4} + K_{y_4 y_2 y_3}\right) + \frac{1}{3} \nabla_{y_3} \left( 3 K_{y_2 y_1 y_4}  - K_{y_1 y_2 y_4}\right)  + \frac{1}{3} \nabla_{y_4} \left( 3 K_{y_2 y_1 y_3} -  K_{y_1 y_2 y_3}\right) + \pi_{y_1 y_2 y_3 y_4},\nonumber\\
&& {}\label{coinc_11} \\
{}[\sigma_{x_1 x_2 x_3 y_4}] &=& - \frac{1}{3}  \nabla_{y_1} \left( K_{y_3 y_2 y_4} + K_{y_4 y_2 y_3}\right) - \frac{1}{3} \nabla_{y_3} \left( 3 K_{y_2 y_1 y_4}  - K_{y_1 y_2 y_4}\right) + \frac{1}{3} \nabla_{y_4} K_{y_1 y_2 y_3} - \pi_{y_1 y_2 y_3 y_4}, \label{coinc_12} \\
{}[\sigma_{x_1 x_2 y_3 y_4}] &=& \frac{1}{3}  \nabla_{y_1} \left( K_{y_4 y_2 y_3} + K_{y_3 y_2 y_4}\right) - \frac{1}{3} \nabla_{y_4} K_{y_1 y_2 y_3}  - \frac{1}{3} \nabla_{y_3} K_{y_1 y_2 y_4} + \pi_{y_1 y_2 y_4 y_3},  \label{coinc_13} \\
{}[\sigma_{x_1 y_2 y_3 y_4}] &=&   - \frac{1}{3} \nabla_{y_1} \left( K_{y_3 y_4 y_2} + K_{y_2 y_4 y_3}\right)   + \frac{1}{3} \nabla_{y_3} K_{y_1 y_4 y_2}  + \frac{1}{3} \nabla_{y_2} K_{y_1 y_4 y_3} +  \nabla_{y_4} K_{y_3 y_1 y_2} - \pi_{y_1 y_4 y_3 y_2},\label{coinc_14}\\ 
{}[\sigma_{y_1 y_2 y_3 y_4}] &=&  \frac{1}{3} \nabla_{y_4} \left( -2 K_{y_2 y_3 y_1} + K_{y_1 y_3 y_2}\right)   - \frac{1}{3} \nabla_{y_2} K_{y_4 y_3 y_1} - \frac{1}{3} \nabla_{y_1} K_{y_4 y_3 y_2} - \nabla_{y_3} K_{y_2 y_4 y_1} + \pi_{y_4 y_3 y_2 y_1}, \label{coinc_15} \\
\pi_{y_1 y_2 y_3 y_4} &:=& \frac{1}{3} \Big[ K_{y_1 y_2}{}^{y} \left( K_{y_3 y_4 y} + K_{y_4 y_3 y} \right) -  K_{y_1 y_3}{}^{y} \left( K_{y_4 y_2 y} + K_{y y_2 y_4} \right) - K_{y_1 y_4}{}^{y} \left( K_{y_3 y_2 y} + K_{y y_2 y_3} \right)  \nonumber \\
&& - 3 K_{y_2 y_1}{}^{y} K_{y_3 y_4 y} + K_{y_3 y_1}{}^{y} K_{y y_2 y_4} + K_{y_4 y_1}{}^{y} K_{y y_2 y_3} + R_{y_1 y_3 y_2 y_4} + R_{y_1 y_4 y_2 y_3} \Big]. \label{pi_def}
\end{eqnarray}
\end{widetext}
Again, we note the added complexity compared to the Riemannian case, in which we have $[\sigma_{x_1 x_2 x_3 x_4}] \stackrel{\rm s}{=} \frac{1}{3} \left(\overline{R}_{y_2 y_4 y_1 y_3} +  \overline{R}_{y_3 y_2 y_1 y_4} \right)$ at the fourth order. In particular, we observe the occurrence of derivatives of the contortion in (\ref{coinc_11})-(\ref{coinc_15}).

Finally, let us collect the basic properties of the so-called parallel propagator $g^{y}{}_{x}:=e^{y}_{(a)} e^{(a)}_{x}$, defined in terms of a parallely propagated tetrad $e^{y}_{(a)}$, which in turn allows for the transport of objects, i.e.\ $V^y=g^y{}_x V^x, \quad  V^{y_1y_2}=g^{y_1}{}_{x_1} g^{y_2}{}_{x_2} V^{x_1x_2}$, etc., along an autoparallel: 
\begin{eqnarray}
g^{y_1}{}_{x} g^{x}{}_{y_2}&=&\delta^{y_1}{}_{y_2}, \quad g^{x_1}{}_{y} g^{y}{}_{x_2}=\delta^{x_1}{}_{x_2}, \label{parallel_1} \\
\sigma^x \nabla_x g^{x_1}{}_{y_1} &=& \sigma^y \nabla_y g^{x_1}{}_{y_1}=0, \nonumber \\
\sigma^x \nabla_x g^{y_1}{}_{x_1} &=& \sigma^y \nabla_y g^{y_1}{}_{x_1}=0, \label{parallel_2} \\
\sigma_x&=&-g^y{}_x \sigma_y, \quad \sigma_y=-g^x{}_y \sigma_x. \label{parallel_3} 
\end{eqnarray}
Note, in particular, the coincidence limits of its derivatives
\begin{eqnarray}
\left[g^{x_0}{}_{y_1} \right] &=& \delta^{y_0}{}_{y_1},  \\
\left[g^{x_0}{}_{y_1 ; x_2} \right] &=& \left[g^{x_0}{}_{y_1 ; y_2} \right] = 0, \label{parallel_4} \\
\left[g^{x_0}{}_{y_1 ; x_2 x_3} \right] &=& - \left[g^{x_0}{}_{y_1 ; x_2 y_3} \right] = \left[g^{x_0}{}_{y_1 ; x_2 x_3} \right] \nonumber \\
&=& - \left[g^{x_0}{}_{y_1 ; y_2 y_3} \right] = \frac{1}{2} R^{y_0}{}_{y_1 y_2 y_3}. \label{parallel_5}
\end{eqnarray}

In the next section we will derive an expanded approximate version of the deviation equation. For this we first work out the expanded version of quantities around the reference world line $Y$. In particular, we make use of the covariant expansion technique \cite{Synge:1960,Poisson:etal:2011} on the basis of the autoparallel world function.

\section{Expanded Riemann-Cartan deviation equation}\label{sec_rc_dev}

For a general bitensor $B_{\dots}$ with a given index structure, we have the following general expansion, up to the third order (in powers of $\sigma^y$):
\begin{eqnarray}
B_{y_1 \dots y_n}&=&A_{y_1 \dots y_n} +  A_{y_1 \dots y_{n+1}} \sigma^{y_{n+1}} \nonumber \\
&& + \frac{1}{2} A_{y_1 \dots y_{n+1} y_{n+2}} \sigma^{y_{n+1}} \sigma^{y_{n+2}} + {\cal O}\left( \sigma^3 \right), \label{expansion_general_yn_1} \\
A_{y_1 \dots y_n}&:=&\left[B_{y_1 \dots y_n}\right] , \label{expansion_general_yn_2} \\
A_{y_1 \dots y_{n+1}}&:=&\left[B_{y_1 \dots y_n ; y_{n+1}}\right] - A_{y_1 \dots y_n ; y_{n+1}} , \label{expansion_general_yn_3} \\
A_{y_1 \dots y_{n+2}}&:=&\left[B_{y_1 \dots y_n ; y_{n+1} y_{n+2}}\right]- A_{y_1 \dots y_n y_0} \left[\sigma^{y_0}{}_{y_{n+1} y_{n+2}}\right] \nonumber \\ 
&&  - A_{y_1 \dots y_n ; y_{n+1} y_{n+2}} - 2 A_{y_1 \dots y_n (y_{n+1} ; y_{n+2})} . \label{expansion_general_yn_4}
\end{eqnarray}
With the help of (\ref{expansion_general_yn_1}) we are able to iteratively expand any bitensor to any order, provided the coincidence limits entering the expansion coefficients can be calculated. The expansion for bitensors with mixed index structure can be obtained from transporting the indices in (\ref{expansion_general_yn_1}) by means of the parallel propagator. 

To develop an approximate form of the generalized deviation equation (\ref{eta_2nd_deriv_alternative_2}) up to the second order, we need the following expansions -- note that we give some explicit intermediate results in appendix \ref{sec_intermediate_details} -- of the derivatives of the world function:
\begin{eqnarray}
\sigma_{y_1 y_2}&=&g_{y_1 y_2} + K_{y_2 y_1 y_3} \sigma^{y_3} + {\cal O}\left( \sigma^2 \right),\label{expansion_explicit_sigma_yy} \\
\sigma_{y_1 x_2}&=&-g_{y_1 x_2} + g_{x_2}{}^y K_{y_3 y y_1} \sigma^{y_3} + {\cal O}\left( \sigma^2 \right),\label{expansion_explicit_sigma_xy} \\
\sigma_{y_1 y_2 y_3}&=&  K_{y_2 y_1 y_3} + \frac{1}{3} \Bigg[ \nabla_{y_4} \left( K_{y_2 y_3 y_1} + K_{y_1 y_3 y_2}\right)  \nonumber \\
&-&  \nabla_{y_2} K_{y_4 y_3 y_1} -  \nabla_{y_1} K_{y_4 y_3 y_2} - 3 \nabla_{y_3} K_{y_2 y_4 y_1} \nonumber \\
&+& 3 \pi_{y_4 y_3 y_2 y_1} \Bigg] \sigma^{y_4} + {\cal O}\left( \sigma^2 \right),  \label{expansion_explicit_sigma_yyy}
\end{eqnarray}
\begin{eqnarray}
\sigma_{y_1 y_2 x_3}&=& g_{x_3}{}^{y_3} \Bigg\{ K_{y_2 y_3 y_1} - \frac{1}{3} \Bigg[ \nabla_{y_3} \left(K_{y_2 y_4 y_1} + K_{y_1 y_4 y_2}\right)  \nonumber \\
&-& \nabla_{y_2} K_{y_3 y_4 y_1} - \nabla_{y_1} K_{y_3 y_4 y_2} \nonumber \\
&+& 3 \pi_{y_3 y_4 y_2 y_1} \Bigg] \sigma^{y_4}\Bigg\} + {\cal O}\left( \sigma^2 \right) , \label{expansion_explicit_sigma_yyx} \\
\sigma_{y_1 x_2 x_3}&=& g_{x_2}{}^{y_2} g_{x_3}{}^{y_3} \Bigg\{ K_{y_3 y_1 y_2} \nonumber \\
&+& \frac{1}{3} \Bigg[ \nabla_{y_2} \left(K_{y_4 y_3 y_1} + K_{y_1 y_3 y_4}\right)  \nonumber \\
&-& \nabla_{y_4} \left( K_{y_2 y_3 y_1} + 3 K_{y_3 y_1 y_2} \right) \nonumber \\
&-& \nabla_{y_1} K_{y_2 y_3 y_4} + 3 \pi_{y_2 y_3 y_4 y_1} \Bigg] \sigma^{y_4} \Bigg\} + {\cal O}\left( \sigma^2 \right). \label{expansion_explicit_sigma_yxx}
\end{eqnarray}
The Jacobi propagators are approximated as
\begin{eqnarray}
H^{x_1}{}_{y_2} & =& g^{x_1}{}_{y_2} + K_{y_3 y_2}{}^{x_1} \sigma^{y_3}+ {\cal O}\left( \sigma^2 \right) , \label{jacobi_prop_1_explicit_xy} \\
 K^{x_1}{}_{y_2}& =& g^{x_1}{}_{y_2} + \left(K_{y_2}{}^{x_1}{}_{y_3}+K_{y_3 y_2}{}^{x_1} \right) \sigma^{y_3} + {\cal O}\left( \sigma^2 \right) , \nonumber \\ \label{jacobi_prop_2_explicit_xy}
\end{eqnarray}
which in turn allows for an expansion of the recurring term entering (\ref{eta_2nd_deriv_alternative_2}):
\begin{eqnarray}
&&\left( K^{x_1}{}_{y_2} u^{y_2} - H^{x_1}{}_{y_2} \frac{D \sigma^{y_2}}{dt} \right)= g^{x_1}{}_{y'} \Bigg[ u^{y'} -  \frac{D \sigma^{y'}}{dt} \nonumber \\
&&+  \left(K_{y_2}{}^{y'}{}_{y_3}+K_{y_3 y_2}{}^{y'} \right) u^{y_2} \sigma^{y_3}  \Bigg]  + {\cal O}\left( \sigma^2 \right) . \label{recurrent_term_explicit}
\end{eqnarray} 

\subsection{Synchronous parametrization}

Before writing down the expanded version of the generalized deviation equation, we will simplify the latter by choosing a proper parametrization of the neighboring curves. The factors with the derivatives of the parameters $t$ and $\tilde{t}$ appear in (\ref{eta_2nd_deriv_alternative_2}) due to the non-synchronous parametrization of the two curves. It is possible to make things simpler by introducing the synchronization of parametrization. Namely, we start by rewriting the velocity as
\begin{equation}
u^y = {\frac {dY^y}{dt}} = {\frac {d\tilde{t}}{dt}} {\frac {dY^y}{d\tilde{t}}}.\label{ut1}
\end{equation}
That is, we now parametrize the position on the first curve by the same variable $\tilde{t}$ that is used on the second curve. Accordingly, we denote 
\begin{equation}
\ut^y = {\frac {dY^y}{d\tilde{t}}}.\label{ut2}
\end{equation} 
By differentiation, we then derive
\begin{eqnarray}
a^y &=& {\frac {d^2\tilde{t}}{dt^2}}\ut^y + \left({\frac {d\tilde{t}}{dt}}\right)^2\at^y,\label{at1}
\end{eqnarray}
where 
\begin{equation}
\at^y = {\frac {D}{d\tilde{t}}}\ut^y = {\frac {D^2Y^y}{d\tilde{t}^2}}.\label{at2}
\end{equation}
Analogously, we derive for the derivative of the deviation vector
\begin{equation}
{\frac {D^2\eta^y}{dt^2}} =  {\frac {d^2\tilde{t}}{dt^2}}{\frac {D\eta^y}{d\tilde{t}}}
+ \left({\frac {d\tilde{t}}{dt}}\right)^2 {\frac {D^2\eta^y}{d\tilde{t}^2}}.\label{Deta2}
\end{equation}
Now everything is synchronous in the sense that both curves are parametrized by $\tilde{t}$.

As a result, the exact deviation equation (\ref{eta_2nd_deriv_alternative_2}) is recast into a simpler form
\begin{eqnarray}
\frac{D^2}{d\tilde{t}^2} \eta^{y_1} &=& - \sigma^{y_1}{}_{y_2} \at^{y_2} - \sigma^{y_1}{}_{x_2} \tilde{a}^{x_2} - \sigma^{y_1}{}_{y_2 y_3} \ut^{y_2} \ut^{y_3} \nonumber \\
&&- 2 \sigma^{y_1}{}_{y_2 x_3} \ut^{y_2} \left( K^{x_3}{}_{y_4} \ut^{y_4} - H^{x_3}{}_{y_4} \frac{D \sigma^{y_4}}{d\tilde{t}} \right) \nonumber \\
&& - \sigma^{y_1}{}_{x_2 x_3}  \left( K^{x_2}{}_{y_4} \ut^{y_4} - H^{x_2}{}_{y_4} \frac{D \sigma^{y_4}}{d\tilde{t}} \right) \nonumber \\
&& \times \left( K^{x_3}{}_{y_5} \ut^{y_5} - H^{x_3}{}_{y_5} \frac{D \sigma^{y_5}}{d\tilde{t}} \right). \label{deveq_master}
\end{eqnarray}

\subsection{Explicit expansion of the deviation equation}

Substituting the expansions (\ref{expansion_explicit_sigma_yy})-(\ref{recurrent_term_explicit}) into (\ref{deveq_master}), we obtain the final result
\begin{widetext}
\begin{eqnarray}
\frac{D^2}{d\tilde{t}^2} \eta^{y_1} &=& \tilde{a}^{y_1} - \at^{y_1} + T_{y_2y_3}{}^{y_1}\ut^{y_2}\frac{D \eta^{y_3}}{d\tilde{t}} 
- \Bigl(K_{y_2y_4}{}^{y_1}\at^{y_2} - K_{y_4y_2}{}^{y_1}\tilde{a}^{y_2} + \Delta^{y_1}{}_{y_2y_3y_4}\ut^{y_2}\ut^{y_3}\Bigr)\,\eta^{y_4} + {\cal O}\left( \sigma^2 \right),\label{final_explicit}
\end{eqnarray}
where we introduced the abbreviation
\begin{eqnarray}
\Delta_{y_1y_2y_3y_4} &:=& 2 \pi_{y_3y_4y_2y_1} - \pi_{y_4y_3y_2y_1} - \pi_{y_2y_3y_4y_1} + T_{y'y_2y_1}T_{y_4y_3}{}^{y'} - 2 \nabla_{y_2}K_{(y_1y_3)y_4} + \nabla_{y_1}K_{y_2y_3y_4} - \nabla_{y_4}K_{y_2y_3y_1}.\label{Delta_explicit}
\end{eqnarray}
\end{widetext}
It should be understood that the last expression is contracted with $\ut^{y_2}\ut^{y_3}$ and hence the symmetrization is naturally imposed on the indices $(y_2y_3)$. 

Equation (\ref{final_explicit}) allows for the comparison of two general world lines in Riemann-Cartan spacetime, which are not necessarily geodetic or autoparallel. It therefore represents the generalization of the deviation equation derived in \cite[(35)]{Puetzfeld:Obukhov:2016:1}.
 
\subsection{Riemannian case}

A great simplification is achieved in a Riemannian background, when
\begin{eqnarray}
  \overline{\Delta}_{y_1y_2y_3y_4} &=& 2 \overline{\pi}_{y_3y_4y_2y_1} - \overline{\pi}_{y_4y_3y_2y_1} - \overline{\pi}_{y_2y_3y_4y_1} \nonumber \\
  &=& \overline{R}_{y_1y_3y_2y_4}, \label{Delta_riem} 
\end{eqnarray}
and (\ref{final_explicit}) is reduced to
\begin{eqnarray}
\frac{D^2}{d\tilde{t}^2} \eta^{y_1} &\stackrel{\rm s}{=}& \tilde{a}^{y_1} - \at^{y_1} - \overline{R}^{y_1}{}_{y_2y_3y_4} \ut^{y_2}\ut^{y_3} \eta^{y_4} + {\cal O}\left( \sigma^2 \right).\nonumber \\ \label{final_riem}
\end{eqnarray}
Along geodesic curves, this equation is further reduced to the well known geodesic deviation (Jacobi) equation. 

\section{Choice of coordinates}\label{sec_coordinate_choice}

To utilize the deviation equation for measurements or in a gravitational compass setup \cite{Synge:1960,Szekeres:1965,Ciufolini:Demianski:1986,Puetzfeld:Obukhov:2016:1}, the occurring covariant total derivatives need to be rewritten and an appropriate coordinate choice needs to be made. The lhs of the deviation equation takes the form:
\begin{eqnarray}
\frac{D^2 \eta_a}{dt^2} &=& \dot{u}^b \nabla_b \eta_a +\stackrel{\circ \circ}{\eta}_a - 2 u^b \Gamma_{ba}{}^d \stackrel{\circ}{\eta}_d - u^b u^c \Gamma_{cb}{}^d \partial_d \eta_{a} \nonumber\\
&& - u^b u^c \eta_e \left(\partial_c\Gamma_{ba}{}^e - \Gamma_{cb}{}^d \Gamma_{da}{}^e - \Gamma_{ca}{}^d \Gamma_{bd}{}^e \right). \nonumber \\ \label{2nd_deriv_expanded}
\end{eqnarray}
Here we used $\stackrel{\circ}{\eta}\!{}^a:=d\eta^a/dt$ for the standard total derivative. 

Observe that the first term on the rhs vanishes in the case of autoparallel curves ($\dot{u}^a:=Du^a/dt=0$). Also note the symmetrization of the connection imposed by the velocities in some terms. 

Rewriting the connection in terms of the contortion and switching to normal coordinates \cite{Veblen:1922,Veblen:Thomas:1923,Thomas:1934,Schouten:1954,Avramidi:1991,Avramidi:1995,Petrov:1969} along the world line, which we assume to be an autoparallel, yields
\begin{eqnarray}
&&\frac{D^2 \eta_a}{dt^2} \stackrel{|_Y}{=} \stackrel{\circ \circ}{\eta}_a + 2 u^b K_{ba}{}^d \stackrel{\circ}{\eta}_d + u^b u^c K_{cb}{}^d \partial_d \eta_{a} \nonumber\\
&& + u^b u^c \eta_e \left(\partial_c K_{ba}{}^e - \frac{2}{3} \overline{R}_{c(ba)}{}^e + K_{cb}{}^d K_{da}{}^e - K_{ca}{}^d K_{bd}{}^e \right). \nonumber \\ \label{2nd_deriv_expanded_autoparallel_normal}
\end{eqnarray}
Note the appearance of a term containing the partial (not ordinary total) derivative of the deviation vector, in contrast to the Riemannian case. 

The first term in the second line may be rewritten as an ordinary total derivative, i.e.\ $u^b u^c \eta_e \partial_c K_{ba}{}^e = u^b \eta_e {\stackrel{\circ}{K}_{ba}{\!\!}^e}$, but this is still inconvenient when recalling the compass equation, which will contain terms with covariant derivatives of the contortion. 

\subsection{Operational interpretation}

At this point, some thoughts about the operational interpretation of the coordinate choice are in order. In particular, it should be stressed that so far we did not specify any {\it physical} theory in which the deviation equation (\ref{deveq_master}) should be applied. Stated the other way round, the derived deviation equation is of completely {\it geometrical} nature, i.e.\ it describes the change of the deviation vector between points on two general curves in Riemann-Cartan spacetime. 

From the mathematical perspective, the choice of coordinates should be solely guided by the simplicity of the resulting equation. In this sense, our previous choice of normal coordinates appears to be appropriate. But what about the physical interpretation or, better, the operational realization of such coordinates? 

Let us recall the coordinate choice in General Relativity in a Riemannian background. In this case, normal coordinates also have a clear operational meaning, which is related to the motion of structureless test bodies in General Relativity. As is well known, such test bodies move along the geodesic equation. In other words, we could -- at least in principle -- identify a normal coordinate system by the local observation of test bodies. If other external forces are absent, normal coordinates will locally\footnote{Here, ``locally'' refers to the observers laboratory on the reference world line. } lead to straight line motion of test bodies. In this sense, there is a clear operational procedure for the realization of normal coordinates.

However, now we are in a more general situation, since we have not yet specified which gravitational theory we are considering in the geometrical Riemann-Cartan background. The physical choice of a gravity theory will be crucial for the operational realization of the coordinates. Recall the form of the equations of motion for a very large class \cite{Puetzfeld:Obukhov:2014:2,Obukhov:Puetzfeld:2015:1} of gravitational theories, which also allow for additional internal degrees of freedom, in particular for spin. In this case, the equations of motion are no longer given by the geodesic equation or, as it is sometimes erroneously postulated in the literature, by the autoparallel equation. In such theories, test bodies exhibit an additional spin-curvature coupling, which leads to non-geodesic motion, even locally. 

How does this impact the operational realization of normal coordinates in such theories? Mainly, one just has to be aware of the fact that for the experimental realization of the normal coordinates, one now has to make sure to use the correct equation of motion and, consequently, the correct type of test body. Taking the example of a theory with spin-curvature coupling, like Einstein-Cartan theory, this would eventually lead to the usage of test bodies with vanishing spin -- since those still move on standard geodesics, and therefore lead to an identical procedure as in the general relativistic case, i.e.\ one adopts coordinates in which the motion of those test bodies becomes rectilinear.  

\section{Conclusions \& Outlook}\label{sec_conclusions}

In this work we investigated the generalization of the deviation equation in a Riemann-Cartan geometry. As a novel technical result, we have developed Synge's world function approach in the non-Riemannian spacetime with curvature and torsion. Our expanded version of the deviation equation (\ref{deveq_master}) can be directly compared to result in the Riemannian context \cite{Puetzfeld:Obukhov:2016:1}. The generalization should serve as a foundation for the test of gravitational theories that make use of post-Riemannian geometrical structures.

As we have discussed in detail, the operational usability of the Riemann-Cartan deviation equation differs from the one in a general relativistic context, which was also noticed quite early in \cite{Hehl:1976}. In particular, it remains to be shown which additional concepts and assumptions are needed in order to fully realize a gravitational compass \cite{Synge:1960,Szekeres:1965,Puetzfeld:Obukhov:2016:1,Ciufolini:Demianski:1986} in a Riemann-Cartan background. In contrast to the Riemannian case, an algebraic realization of a gravitational compass on the basis of the deviation equation is out of the question due to the appearance of derivatives of the torsion even at the lowest orders.   

An interesting question for future works is the possible application of (\ref{deveq_master}) to the analysis of motion of (micro)structured test bodies. In particular, it seems worthwhile to search for new ways to map the gravitational field with the help of such a deviation equation and work out its implications for various applications, aiming for novel tests of relativistic gravity theories.  

\section*{Acknowledgements}
This work was supported by the Deutsche Forschungsgemeinschaft (DFG) through the grant PU 461/1-1 (D.P.). The work of Y.N.O. was partially supported by PIER (``Partnership for Innovation, Education and Research'' between DESY and Universit\"at Hamburg) and by the Russian Foundation for Basic Research (Grant No. 16-02-00844-A). 

\appendix

\section{Notation \& Conventions}\label{sec_notation}

\begin{table}
\caption{\label{tab_symbols}Directory of symbols.}
\begin{ruledtabular}
\begin{tabular}{ll}
Symbol & Explanation\\
\hline
&\\
\hline
\multicolumn{2}{l}{{Geometrical quantities}}\\
\hline
$g_{a b}$ & Metric\\
$\delta^a_b$ & Kronecker symbol \\
$x^{a}, y^a$ & Coordinates \\
$\Gamma_{a b}{}^c$ & Connection \\
$\overline{\Gamma}_{a b}{}^c$ & Levi-Civita connection \\
$R_{a b c}{}^d$ & Curvature \\
$T_{ab}{}^c$ & Torsion\\
$K_{ab}{}^c$ & Contortion\\
$\sigma$ & World function\\
$\eta^y$ & Deviation vector\\
$g^{y_0}{}_{x_0}$ & Parallel propagator\\
&\\
\hline
\multicolumn{2}{l}{{Misc}}\\
\hline
$Y(t), X(\tilde{t})$ & (Reference) world line\\
$u^a$ & Velocity \\
$a^b$ & Acceleration \\
$K^{x}{}_{y}, H^{x}{}_{y}$ & Jacobi propagators \\
$A_{y_1 \dots y_n}$ & Expansion coefficient \\
$\pi_{y_1 y_2 y_3 y_4}$ & Auxiliary quantities \\
&\\
\hline
\multicolumn{2}{l}{{Operators}}\\
\hline
$\partial_i$, ``${\phantom{a}}_{,}$'' & Partial derivative \\
$\nabla_i$, ``${\phantom{a}}_{;}$'' & Covariant derivative \\ 
$\frac{D}{dt} = $``$\dot{\phantom{a}}$'' & Total cov. derivative \\
$\frac{d}{dt} = $``$\stackrel{\circ}{\phantom{a}}$'' & Total  derivative \\
``$[ \dots ]$''& Coincidence limit\\
``$\overline{\phantom{A}}$''& Riemannian object\\
&\\
\end{tabular}
\end{ruledtabular}
\end{table}

The curvature and the torsion are defined w.r.t.\ the general connection $\Gamma_{ab}{}^c$ as follows:
\begin{eqnarray}
R_{abc}{}^d &:=& \partial_a \Gamma_{bc}{}^d - \partial_b \Gamma_{ac}{}^d + \Gamma_{an}{}^d \Gamma_{bc}{}^n - \Gamma_{bn}{}^d \Gamma_{ac}{}^n, \nonumber \\ \label{curvature_def} \\
T_{ab}{}^c &:=& \Gamma_{ab}{}^c -  \Gamma_{ba}{}^c. \label{torsion_def} 
\end{eqnarray}
The symmetric Levi-Civita connection $\overline{\Gamma}_{kj}{}^i$, as well as all other Riemannian quantities, are denoted by an additional overline. For a general tensor $A$ of rank $(n,l)$ the commutator of the covariant derivative thus takes the form:
\begin{eqnarray}
\left(\nabla_a \nabla_b - \nabla_b \nabla_a \right) A^{c_1 \dots c_n}{}_{d_1 \dots d_l} = - T_{ab}{}^e \nabla_e A^{c_1 \dots c_k}{}_{d_1 \dots d_l} \nonumber \\
+ \sum^k_{i=1} R_{abe}{}^{c_i} A^{c_1 \dots e \dots c_k}{}_{d_1 \dots d_l} - \sum^l_{j=1} R_{abd_j}{}^{e} A^{c_1 \dots c_k}{}_{d_1 \dots e \dots d_l}. \nonumber \\\label{cov_deriv}
\end{eqnarray}
In addition to the torsion, we define the contortion $K_{kj}{}^i$ with the following properties
\begin{eqnarray}
K_{kj}{}^i&:=& \overline{\Gamma}_{kj}{}^i - \Gamma_{kj}{}^i, \label{distorsion_def}\\
K_{kji}&=& - \frac{1}{2} \left( T_{kji} + T_{ikj} + T_{ijk} \right) , \label{torsion_distorsion_1}\\
T_{kj}{}^i&=&-2K_{[kj]}{}^i. \label{torsion_distorsion_2}
\end{eqnarray}
The signature of the spacetime metric is assumed to be $(+1,-1,-1,-1)$. 

As usual, $\sigma^y{}_{x_1\dots y_2\dots}:=\nabla_{x_1}\dots\nabla_{y_2}\dots (\sigma^y)$ denote the higher-order covariant derivatives of the world function.

\section{Intermediate results}\label{sec_intermediate_details}

Here we give some intermediate results of the derivation of the expansions the world function derivatives around the reference world line $Y$: 

\begin{widetext}
\begin{eqnarray}
\sigma_{y_1 y_2}&=& g_{y_1 y_2} + [\sigma_{y_1 y_2 y_3}] \sigma^{y_3} + \left(\frac{1}{2} [\sigma_{y_1 y_2 y_3 y_4}] - [\sigma_{y_1 y_2 y_3}]_{;y_4} - \frac{1}{2} [\sigma_{y_1 y_2 y_5}] [\sigma^{y_5}{}_{y_3 y_4}] \right) \sigma^{y_3} \sigma^{y_4} + {\cal O}\left( \sigma^3 \right) , \label{expansion_sigma_yy} \\
\sigma_{y_1 y_2 y_3}&=& [\sigma_{y_1 y_2 y_3}] + \bigg([\sigma_{y_1 y_2 y_3 y_4}] - [\sigma_{y_1 y_2 y_3}]_{;y_4} \bigg) \sigma^{y_4} +\frac{1}{2} \bigg( [\sigma_{y_1 y_2 y_3 y_4 y_5}] - [\sigma_{y_1 y_2 y_3}]_{; y_4 y_5} -  2 [\sigma_{y_1 y_2 y_3 y_4}]_{; y_5} \nonumber \\
&& - [\sigma_{y_1 y_2 y_3 y_0}] [\sigma^{y_0}{}_{y_4 y_5}]  + [\sigma_{y_1 y_2 y_3}]_{;y_0} [\sigma^{y_0}{}_{y_4 y_5}] \bigg) \sigma^{y_4} \sigma^{y_5} + {\cal O}\left( \sigma^3 \right) , \label{expansion_sigma_yyy}\\
\sigma_{y_1 x_2}&=& -g_{y_1 x_2} + g_{x_2}{}^{y} [\sigma_{y_1 x y_3}] \sigma^{y_3} + \frac{1}{2} \bigg( g_{x_2}{}^{y} [\sigma_{y_1 x y_3 y_4}] - g_{x_2}{}^{y_2} g_{y_1 y} \left[g_{y_2}{}^{x}{}_{;y_3 y_4}\right] \nonumber \\
&& - 2 g_{x_2}{}^{y} [\sigma_{y_1 x (y_3}]_{;y_4)} - g_{x_2}{}^{y} [\sigma_{y_1 x y_5}] [\sigma^{y_5}{}_{y_3 y_4}] \bigg) \sigma^{y_3} \sigma^{y_4}  + {\cal O}\left( \sigma^3 \right) . \label{expansion_sigma_yx} 
\end{eqnarray} 
\end{widetext}

\bibliographystyle{unsrtnat}
\bibliography{devrc_bibliography}
\end{document}